\documentclass[twocolumn,pre]{revtex4-1}
\usepackage{graphics,graphicx,epsfig}
\usepackage{color}
\usepackage{ulem}

\begin{document}

\title{Crossover from compact to branched films in electrodeposition with surface diffusion}

\author{F. D. A. Aar\~ao Reis}
\email{reis@if.uff.br}
\affiliation{Instituto de F\'{i}sica, Universidade Federal Fluminense,
Avenida Litor\^{a}nea s/n, 24210-340 Niter\'{o}i, RJ, Brazil}
\author{Dung di Caprio}
\email{dung.di-caprio@chimie-paristech.fr}
\affiliation{PSL Research University, Chimie ParisTech - CNRS, Institut de Recherche
de Chimie Paris, 75005, Paris, France}
\author{Abdelhafed Taleb}
\email{abdelhafed.taleb@upmc.fr}
\affiliation{Universit\'{e} Pierre et Marie Curie, 4 place Jussieu, 75231, Paris, France}

\begin{abstract}
We study a model for thin film electrodeposition in which instability development by
preferential adsorption and reduction of cations at surface peaks competes with surface
relaxation by diffusion of the adsorbates.
The model considers cations moving in a supported electrolyte, adsorption and reduction
when they reach the film surface, and consequent production of mobile particles
that execute activated surface diffusion, which is represented by a sequence of random hops
to neighboring lattice sites with a maximum of $G$ hop attempts ($G\gg 1$), a detachment
probability $\epsilon <1$ per neighboring particle, and a no-desorption condition.
Computer simulations show the formation of a compact wetting layer
followed by the growth of branched deposits.
The maximal thickness $z_c$ of that layer increases with $G$, but is weakly affected by
$\epsilon$.
A scaling approach describes the crossover from smooth film growth to unstable growth and
predicts $z_c\sim G^{\gamma}$, with $\gamma = 1/{\left[ 2\left( 1-\nu\right)\right]}\approx 0.43$,
where $\nu\approx 0.30$ is the inverse of the dynamical exponent of the Villain-Lai-Das Sarma
equation that describes the initial roughening.
Using previous results for related deposition models, the thickness $z_c$ can be predicted as a
function of an activation energy for terrace surface diffusion and the temperature,
and the small effects of the parameter $\epsilon$ are justified.
These predictions are confirmed by the numerical results with good accuracy.
We discuss possible applications, with a particular focus on the growth of multifuncional
structures with stacking layers of different porosity.
\end{abstract}

\maketitle

\section{Introduction}
\label{intro}

Electrodeposition is a relatively low cost technique that allows the production
of multifunctional thin films and other nanostructures with various structural, optical,
electrical, and magnetic properties \cite{Gamburg2011,oja}.
Electrodeposition is strongly affected by the diffusive mass transport of species from
the electrolyte to the growing surface,
which may lead to morphological instabilities \cite{aogaki,elezgaray,haataja}.
This feature motivated the proposal of several stochastic models, starting from the
diffusion-limited aggregation (DLA) model \cite{DLA,meakin1983,argoul1988}.
Other models consider mechanisms for controlling the diffusional instability
and analyze their effects on the long-time morphology of the electrodeposited films
\cite{haataja,uwaha,castro2000,nicoliJSTAT2009,nielsen}.
On the other hand, several recent works show that the mass transport of adsorbed atoms
or molecules on the film surface also affects its morphology
\cite{grunder,PB,aryanfar,brandt2015,mahboob} and detailed investigations of the first stages
of electrodeposition of various materials reveal the microscopic mechanisms of adsorbate
diffusion \cite{ustarroz,desai,zimmer}.
Analytic and simulation models also account for those processes
\cite{aogaki,PB,guoEA2010,tanyakarn,mamme}.

The interplay between the diffusional limitations in solution and the surface relaxation
(via adsorbate diffusion) is important to understand the nucleation
and growth stages and may have impact on the production of
films and other nanostructures with the desired properties.
The production of multifunctional films is possible, for instance, by controlled
changes in the film structure and architecture \cite{hafed03}.
In many cases, deposited films show different stacking layers with different porosities and
thicknesses \cite{hafed07}, whose different physical and/or chemical properties
ensure multifunctional behavior.
For example, regarding photovoltaic applications, a layer with large branches and reduced
specific surface area ensures the light harvesting, while a layer with small branches enhances
the specific surface and favors loading of large molecules (e. g. dyes).
These different functions may correspond to conflicting requirements, thus it is necessary to
understand how to control the thickness of different layers and the transitions between them.  
This transition is governed by parameters such as temperature, concentration, potential
\cite{talebJMCA2013,hafed08}, substrate properties \cite{hafed09},
and deposition method \cite{hafed10}.

In this work, we introduce a stochastic model for electrochemical deposition that represents
cation diffusion in a supported electrolyte and diffusion of the atoms or
molecules adsorbed on the electrode or on the film surface.
Using numerical simulations, we show the formation of a compact layer near the electrode
(a wetting layer) and the development of a branched morphology after a characteristic
crossover time.
Since these are typical features that provide multifunctional behavior,
our aim is to understand how physico-chemical parameters control the length and
time scales involved in that crossover.
A scaling approach and some results from related models are used to predict the relation between
the crossover time and the model parameter related to the
surface diffusion coefficient (which in turn depends on an activation
energy and temperature).
These predictions are in good agreement with the numerical results.

Images of electrodeposited films of different metals have already shown the formation of
compact deposits with relatively smooth surfaces before the development of branched morphology
\cite{argoul1988,kahanda,pastor,lafouresse,debruyn,legerPRE1998,pasquale}.
However, those works did not analyze this crossover in detail because their focus were
long time growth properties, such as surface instability and anomalous scaling of the roughness.
Most theoretical approaches for electrodeposition have the same focus;
for instance, in systems with instability control mechanisms, a transition from initial
unstable growth to standard kinetic roughening may be observed \cite{castro2000,nicoliJSTAT2009}.
Instead, the short time crossover analyzed here occurs in the opposite direction: 
from initial kinetic roughening dominated by surface diffusion \cite{villain,laidassarma}
to long time instability developement.

The rest of this paper is organized as follows.
In Section \ref{basics}, we present the electrodeposition model and the quantities of
interest in this work.
In Section \ref{results}, we present simulation results for the film density profiles and
for the initial kinetic roughening.
In Sec. \ref{relation}, we analyze the relations with other deposition models with adatom surface
diffusion.
In Section \ref{crossover}, a scaling approach is used to determine the film thickness at the
crossover from compact to branched morphology, relations to microscopic parameters are
analyzed, and possible applications are discussed.
In Section \ref{conclusion}, a summary of results and conclusions is presented.

\section{Model and basic concepts}
\label{basics}

\subsection{Model definition}
\label{defmodel}

This subsection is intended to the presentation of the stochastic rules of the model.
The physical interpretation of the model is discussed
in Sec. \ref{interpretation}.

The model is defined in a simple cubic lattice with lattice constant $a$ and size $La$
in the $x$ and $y$ directions, with periodic boundary conditions in those directions.
The substrate (working electrode) is at the plane $z=0$ and
lattice sites with $z>0$ initially contain a supported electrolyte.

The deposition occurs by sequential incorporation of aggregated particles,
each one occupying one lattice site.
These particles may represent atoms or molecules, depending on the electrochemical reaction
of interest, which means that the lattice constant $a$ is of the order of some tenths of a nanometer.
The incorporation of each particle occurs after a sequence of steps:
diffusion of a cation in solution, adsorption on the substrate or on the surface of
the growing deposit, reduction and consequent formation of a mobile particle, and
surface diffusion of this particle until it reaches a final aggregation position,
where it will be called an aggregated particle.
This sequence of steps is described below.

\begin{figure}[!ht]
\center
\includegraphics[clip,width=.5\textwidth,angle=0]{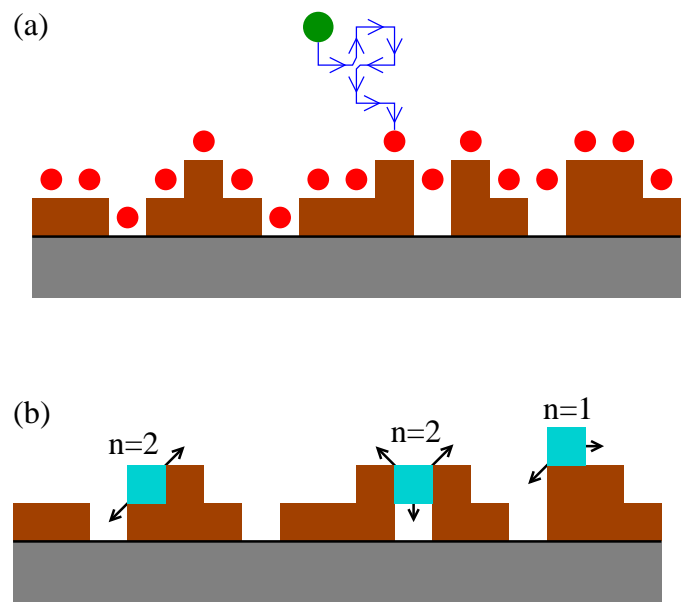}
\caption{Two-dimensional representation of the deposition model.
(a) The region near the substrate (grey) with some aggregated particles (brown squares)
has points available for cation adsorption and reduction indicated by red circles.
A cation (green circle) executes a random walk in solution (blue line with arrows) until
it reaches one of those points.
(b) Possible hops of mobile particles (light blue squares) at three different
positions with the corresponding number of NNs indicated.
}
\label{model}
\end{figure}

A two-dimensional illustration of the diffusion of the cation in the solution until the
adsorption-reduction point is shown in Fig. \ref{model}(a).
First, a position $\left( x,y,h_{max}+45a\right)$ is chosen for releasing the cation, with
random $x$ and $y$ and where $h_{max}$ is the maximal value of $z$ of a previously aggregated
particle.
This choice describes an approximately uniform
distribution of cations at a height which is not very close to the deposit.
The cation executes random walks to nearest neighbor (NN) sites in the electrolyte.
Its position $z$ cannot exceed $h_{max}+45a$, i. e. it is reflected at that height.
When the cation reaches a site in which at least one NN is an aggregated particle
or a substrate site, it is immediately reduced.
The product of the electrochemical reaction occupies the same site and
is hereafter called a mobile particle.
No activation barrier is considered for the processes of cation adsorption and reduction.

During the growth of a film, the mobility of an adsorbed species is expected to increase
with temperature and to depend on the local surface morphology.
The diffusion length of this species will be finite because it will be eventually covered
(buried) by other atoms or molecules.
For these reasons, here the mobile particle executes a sequence of hops to neighboring sites,
with rules that privilege (but do not obly) the aggregation at points with high
coordination, and the number of hops is constrained to have a maximal value.
Moreover, our model assumes that the adsorption states have energy sufficiently small to forbid
desorption, even in the cases where large diffusion lengths are considered.

The diffusion of the mobile particle is represented by a sequence of $G$  attempts to
hop to neighboring sites, beginning at the point in which that particle was produced.
In each attempt, there are two conditions for the hop to be executed:

\par\noindent A) Let $n$ be the number of occupied NN sites of the current position of the mobile
particle (occupied sites are those with an aggregated particle or in the substrate) and
\begin{equation}
P_{hop} = \epsilon^{n-1} ,
\label{Pstep}
\end{equation}
with
\begin{equation}
\epsilon=\exp{\left( -E_b/k_BT\right)} ,
\label{Eb}
\end{equation}
where $E_b>0$ is an activation energy and $T$ is the temperature (consequently $\epsilon <1$).
With probability $1-P_{hop}$, the hop is not executed; with probability $P_{hop}$, the
second condition is checked.

\par\noindent B) A target site is randomly chosen among the NN and next nearest neighbor (NNN)
sites of the current position.
The hop is executed if the target site is empty and has at least one NN which is occupied;
otherwise, the hop is not executed.
This condition rules out desorption of the mobile particle. 

\par\noindent After $G$ attempts to hop, the mobile particle becomes an aggregated
particle at its final position.

Fig. \ref{model}(b) shows the allowed directions for a hop of three particles at different
positions of the deposit in a two dimensional representation of the model.

The time $\tau$ is defined as the time of aggregation of $L^2$ particles, i. e. the
time for adsorption of one monolayer.
The coverage at time $t$ is $t/\tau$, which corresponds to the number of monolayers in the case
of a compact deposit.
We will assume that $\tau$ is constant, which means that the mass of the deposit increases
linearly with the time $t$;
this assumption approximately corresponds to galvanostatic deposition.

Simulations of the model were performed in lattices with $L=1024$ until 
the maximal height reached the value $h_{max}=2000a$.
Several values of $G$ were considered, from ${10}^2$ to $5\times {10}^4$.
The values of $\epsilon$ were between $0.05$ and $0.25$.
For each set of parameters, $9$ different realizations were generated.
The lattices are very large, thus each realization contains a very large number of
different microscopic environments and the average quantities have relatively small
fluctuations (except after the development of long branches, but these long time
features are not analyzed in this work).
Simulations for some model parameters were performed in lattices with $L=1536$
for checking for finite-size effects, but maximal heights had to be smaller due
to computational limitations.

\subsection{Interpretation of model parameters}
\label{interpretation}

The diffusive motion of cations in solution is expected in a supported electrolyte, since the
addition of a high concentration of inert ions significantly increases the conductivity,
reducing the electric field and suppressing migration of the electroactive ions;
for details, see Ref. \protect\cite{wangbook}.
At long deposition times, cation migration due to the electric field is expected to affect the
film features, as discussed in Sec. \ref{relationmodels}, but these effects are not considered
in the present model.

For any set of model parameters, the formation of the mobile particle when the cation
has an occupied NN means that the sticking coefficient for the cation is $1$,
which corresponds to the absence of activation barriers to go from solution to an
adsorption state.
Moreover, the cation reduction at the point of adsorption means that the rate of the
electrochemical reaction is very large,
which is possible if the cathodic potential is sufficiently negative.

Now consider the case $G=0$.
The product of the electrochemical reaction is not mobile in this case, i. e. it permanently
aggregates at the position where the cation was reduced.
Thus, the model becomes equivalent to the diffusion-limited aggregation (DLA) of
Witten and Sander \cite{DLA}, which leads to the formation of highly branched fractal deposits;
see e. g. Ref. \protect\cite{meakin1983} for the case of growth on a surface.
For small $G$, we also expect to obtain film features similar to those of DLA.

Now consider the growth of a real film in which the adsorbed species is mobile.
We expect that this species has a local diffusion coefficient ${\cal D}$ which increases
with temperature and which depends on the local surface morphology; typically, as the number of
neighbors increase, the local value of ${\cal D}$ decreases.
In a film terrace, the diffusion coefficient is expected to have its maximal possible value $Da^2$,
where $D$ has an Arrhenius form
\begin{equation}
D = h_0 \exp{\left( -E_s/k_BT\right)} ,
\label{defD}
\end{equation}
where $E_s$ is a terrace activation energy and $h_0$ is a hopping frequency.
On the other hand, the diffusion process will take place in a finite time because the mobile species
will be eventually covered (buried) by other atoms or molecules.
This time decreases as the deposition rate $F$ increases, with $F$ defined
as the number of reduced cations per substrate site per unit time.

Our model is consistent with these features, although it simplifies the dynamics of adsorbed
particles.
The mobile particle in a film terrace or in the substrate ($n=1$) executes $G$ hops, i. e.
the surface diffusion always takes place in a finite time.
Moreover, if the mobile particle has more than one neighbor ($n>1$), then the probability $P_{hop}$
[Eq. (\ref{Pstep})] represents the reduction in the local value of ${\cal D}$ in comparison
with the terrace value $D$.

For the above reasons, we also expect that $G$ increases with $D$ and decreases with $F$.
The discussion on the relation between $G$ and the ratio $D/F$ is postponed to 
Sec. \ref{relation} because it will be motivated by the numerical results and requires the
presentation of previous results for other deposition models \cite{CDLM,cv}.
However, we anticipate that $G$ is related to $D/F$ by a power law, thus Eq. (\ref{defD})
implies that $G$ depends on the temperature in an Arrhenius form.

Also note that the dependence of $P_{hop}$ on $T$ [Eqs. (\ref{Pstep}) and (\ref{Eb})] is the same
of the Clarke-Vvedensky (CV) model \cite{cv,etb} of thin film deposition, in which the bond energy
$E_b$ per NN describes the effect of the local surface morphology.
The present model and the CV model privilege the aggregation at points with high coordination
because $P_{hop}$ is very small at those points.

Similarly to many other thin film growth models \cite{etb,barabasi}, our model assumes that the
adsorbed states have energy which is sufficiently small compared to
that in solution, thus desorption is forbidden.
This occurs even in the cases where small surface diffusion barriers (and large diffusion lengths)
are considered.

Finally, we recall that the use of limited mobility models (such as the model proposed here) instead
of collective diffusion models (e. g. CV model) is essential for the simulation of large deposits
in a broad range of parameters; see e. g. Ref. \protect\cite{CDLM}.

\subsection{Definition of basic quantities}
\label{quantities}

The density $\rho$ was calculated at each level $z>0$ for several times.
It is given by the ratio of the number of aggregated particles at that level and the
horizontal cross-sectional area $L^2$.
The set of deposited particles with position $\left( x,y\right)$ will be called a column
of the film.

Since the deposit has overhangs and pores, the height of a given column, $h\left( x,y\right)$,
is defined as the largest position $z$ of a particle at that column.
The set $\{ h\left( x,y\right)\}$ defines the outer surface of the deposit.

The local roughness in a square box of lateral size $r$ at time $t$ is defined as
\begin{equation}
w\left( r,t\right)\equiv {\langle {\overline{{\left( h-\overline{h}\right)}^2}}^{1/2}\rangle} ,
\label{defw}
\end{equation}
where the overbars denote a spatial average in each box position
(root-mean square height
fluctuation inside the square box at a given position) and the angular brackets denote
a configurational average over all positions of the box (which glides parallel to the
substrate) and over different configurations of the deposit at time $t$.

The autocorrelation function \cite{zhao,siniscalco} is defined as
\begin{equation}
\Gamma\left( s,t\right) \equiv \frac{ \left\langle {\left[ \tilde{h}\left( {\vec{r}}_0+\vec{s},t\right)
\tilde{h}\left( {\vec{r}}_0,t\right) \right]}^2 \right\rangle }{W^2} ,\qquad s\equiv |\vec{s}| ,
\qquad \tilde{h}\equiv h-\overline{h},
\label{defcorr}
\end{equation}
where $\overline{h}$ is the global average height of
the deposit at time $t$ and $W^2$ is the global square roughness
[defined as in Eq. (\ref{defw}) for $r=L$].
The configurational average in Eq. (\ref{defcorr}) is taken over different initial positions
${\vec{r}}_0$, different orientations of $\vec{s}$ (directions $x$ and $y$), and different deposits.
Note that $\Gamma\left( 0,t\right) =1$ at any time $t$.

The best method to estimate the lateral correlation length from the autocorrelation
function depends on details of the interface morphology \cite{zhao,siniscalco}.
In surfaces with some patterned structure (e. g. mounds), $\Gamma$ oscillates from positive
to negative values as $s$ increases, at fixed time; in these cases, the correlation length
$\xi\left( t\right)$ may be defined as the first zero of $\Gamma\left( s,t\right)$.
In surfaces with no pattern, $\Gamma\left( s,t\right)$ may oscillate with $s$ before
crossing the value $\Gamma =0$, so that a more suitable definition of a correlation length
is \cite{lrd2015}
\begin{equation}
\Gamma\left( \xi,t\right) = k ,
\label{defxi1}
\end{equation}
with some constant $0<k<1$. 
In this work, we calculate $\xi$ using Eq. (\ref{defxi1}) in forms consistent with both
approaches: $k=0$ (first zero of $\Gamma$) and $k=0.1$.

\subsection{Kinetic roughening}
\label{roughening}

In systems with normal roughening, the expected scaling of the local roughness in
large substrates is \cite{barabasi,fv}
\begin{equation}
w\left( r,t\right) = r^{\alpha} f{\left( \frac{r}{\xi}\right)} ,
\label{fvlocal}
\end{equation}
where $\alpha$ is the roughness exponent and $f$ is a scaling function.
For $x\equiv r/\xi \ll 1$ (small box sizes), $f\left( x\right)$ is constant and we obtain
\begin{equation}
w \sim r^{\alpha} .
\label{defbeta}
\end{equation}
This means that a single curve with slope $\alpha$ is observed in $\log{w}\times\log{r}$ plots
for different times.
For large $r$, those curves split because the saturation value of $w$ (the global roughness)
depends on time.
The correlation length is expected to scale as
\begin{equation}
\xi \sim t^{\nu} ,
\label{xiscaling}
\end{equation}
where $\nu$ is the inverse of the dynamical exponent, which determines the rate of lateral
propagation of height fluctuations.
Equation (\ref{xiscaling}) is also the scaling of the correlation length $\xi_1$ defined in
Eq. (\ref{defxi1}).

In systems with intrinsic anomalous roughening \cite{ramasco}, the local roughness scales as
\begin{equation}
w\left( r,t\right) = r^{\alpha_{loc}} t^\kappa g{\left( \frac{r}{\xi}\right)} ,
\label{fvanomalous}
\end{equation}
where $\alpha_{loc}$ is the local roughness exponent, the exponent $\kappa >0$ represents the degree
of anomaly of the system, and the scaling function $g\left( x\right)$ is also constant for $x\ll 1$.
The curves $\log{w}\times\log{r}$ at different times now split for small $r$.
The exponent $\kappa$ is usually called local growth exponent ($\beta_{loc}$)
in experimental works \cite{lafouresse,huo}.

When the roughening is dominated by surface diffusion of the adsorbed species,
the evolution of the film surface in the hydrodynamic limit (large distances, long times)
is expected to be described by the Villain-Lai-Das Sarma (VLDS)
\cite{villain,laidassarma} equation:
\begin{equation}
{{\partial H}\over{\partial t}} = -\nu_4{\nabla}^4 H +
\lambda_{4} {\nabla}^2 {\left( \nabla H\right) }^2 + \eta (\vec{r},t) ,
\label{vlds}
\end{equation}
where $H\left( \vec{r},t\right)$ is a coarse-grained height variable, $\nu_4$ and $\lambda_{4}$
are constants, and $\eta$ is a Gaussian white noise.

In VLDS growth in two-dimensional substrates, the best estimates of scaling exponents
are obtained from simulations of the conserved restricted solid-on-solid models
\cite{crsosreis,carrasco2016}, and are very close to one-loop renormalization values \cite{janssen}:
$\alpha \approx 0.67$, $\nu\approx 0.30$, and $\beta\approx 0.20$.

\section{Simulation results}
\label{results}

\subsection{Basic features of the deposits}
\label{deposits}

\begin{figure}[!ht]
\includegraphics[clip,width=.45\textwidth,angle=0]{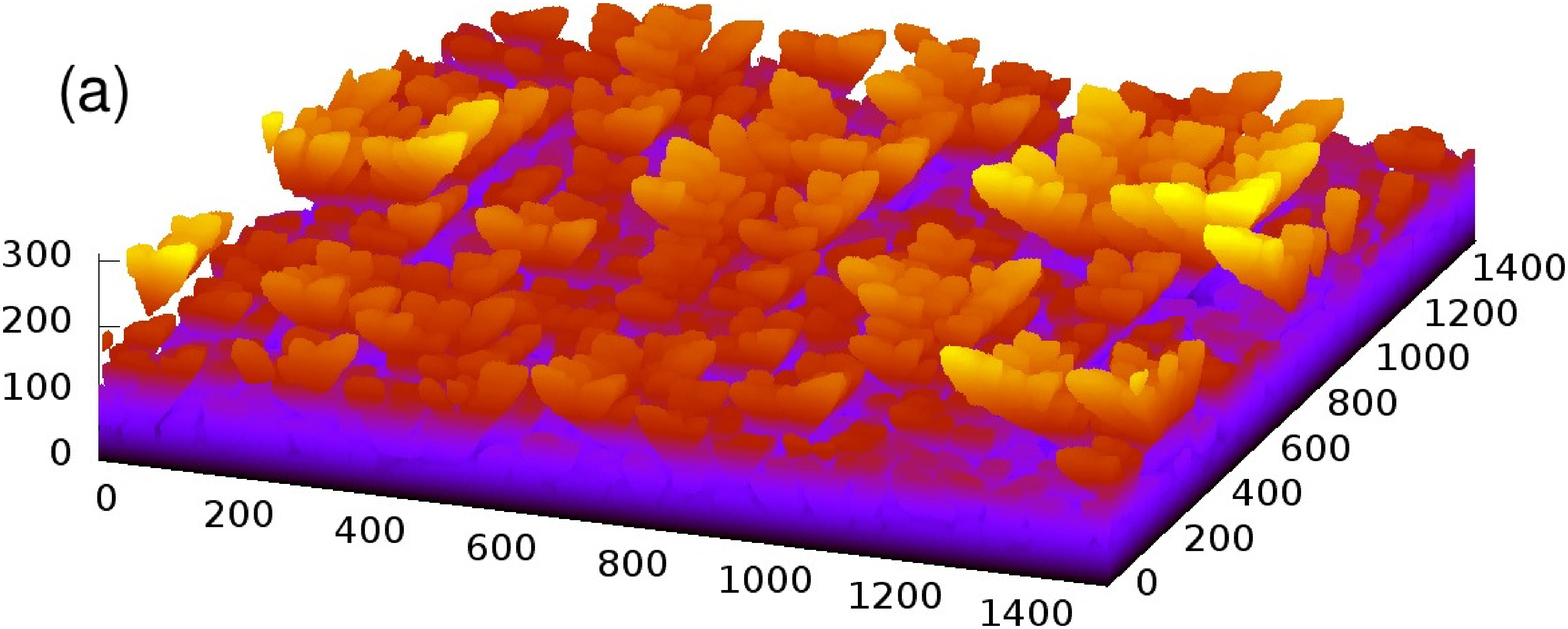}\\
\includegraphics[clip,width=.45\textwidth,angle=0]{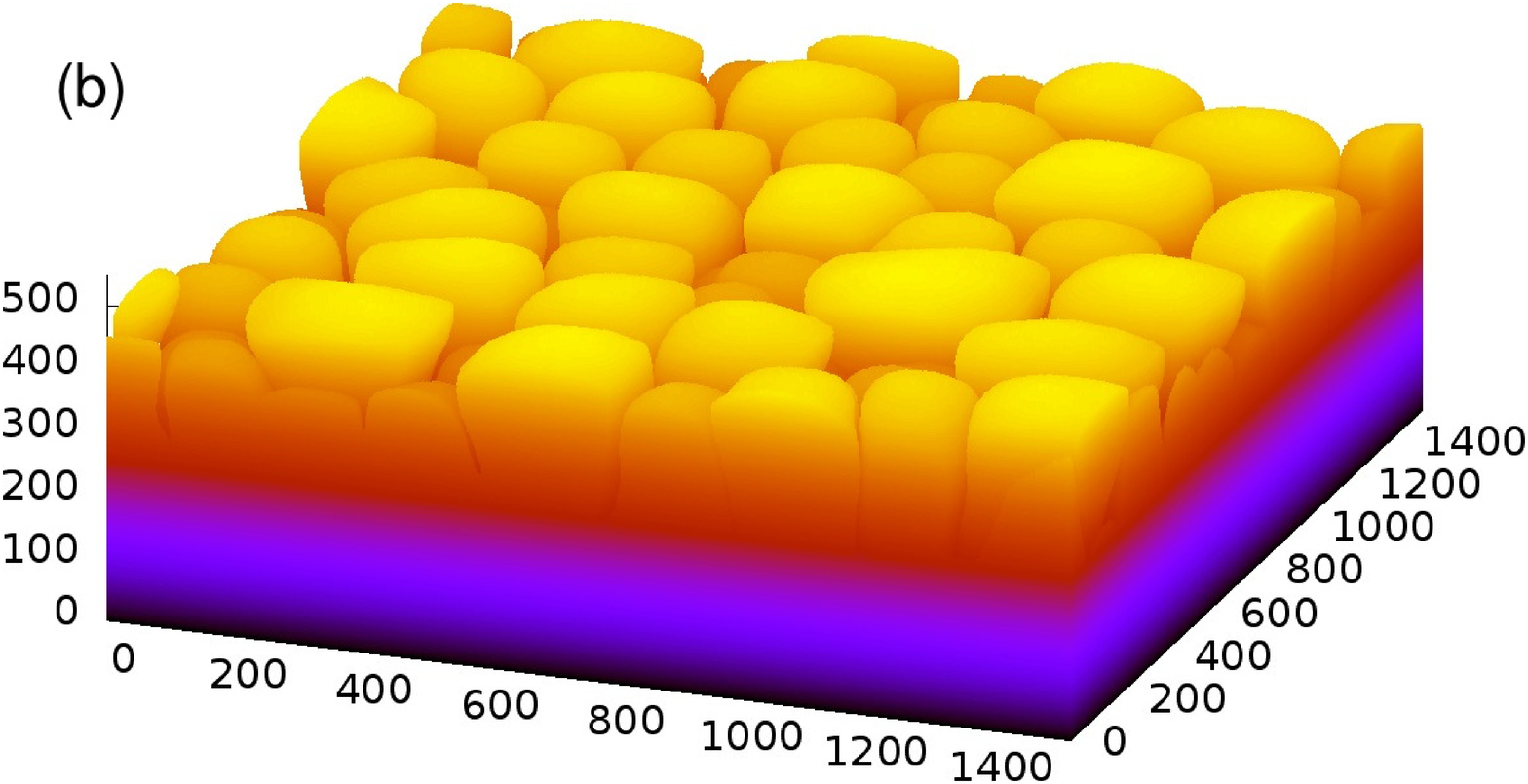}
\caption{
Deposits grown for (a) $G=2000$ and (b) $G=50000$, both with $\epsilon =0.1$,
on a $1536\times 1536$ substrate with periodic boundary conditions in
horizontal directions.
}
\label{sections}
\end{figure}

In Figs. \ref{sections}a and \ref{sections}b, we show films grown with $G=2\times{10}^3$ and
$G=5\times{10}^4$, respectively, both with $\epsilon = 0.1$.

For $G=2\times{10}^3$, a thin wetting layer is formed at short times
[darkest colors in Fig. \ref{sections}(a)].
When some hills and valleys appear at the film surface, the diffusive cation flux
leads to preferential growth at those hills.
This leads to growth of rounded mounds and eventual formation of overhangs; this is the case
of bumps separated by narrow gaps at $z\approx 100$ and below in Fig. \ref{sections}(a).
When high branches are formed [lightest colors in Fig. \ref{sections}(a)], they reduce all
diffusing cations and the mobile particles cannot reach the wetting layer.
Thus, the thickness of the wetting layer remains approximately constant while the
highest branches continue to grow.

For $G=5\times{10}^4$, the film is compact until a thickness $h\approx 200a$ is reached,
as shown in Fig. \ref{sections}(b) (darkest layers).
It also has a relatively smooth surface in these conditions.
The instability is also developed when thick branches separated by narrow gaps begin to grow
[lightest colors in Fig. \ref{sections}(b)]; their sizes are much larger than those in Fig.
\ref{sections}(a).
Again, the wetting layer stops growing when the branches appear because the mobile particles
aggregate before reaching that layer.

\subsection{Density profile and crossover thickness}
\label{densityprofile}

\begin{figure}[!ht]
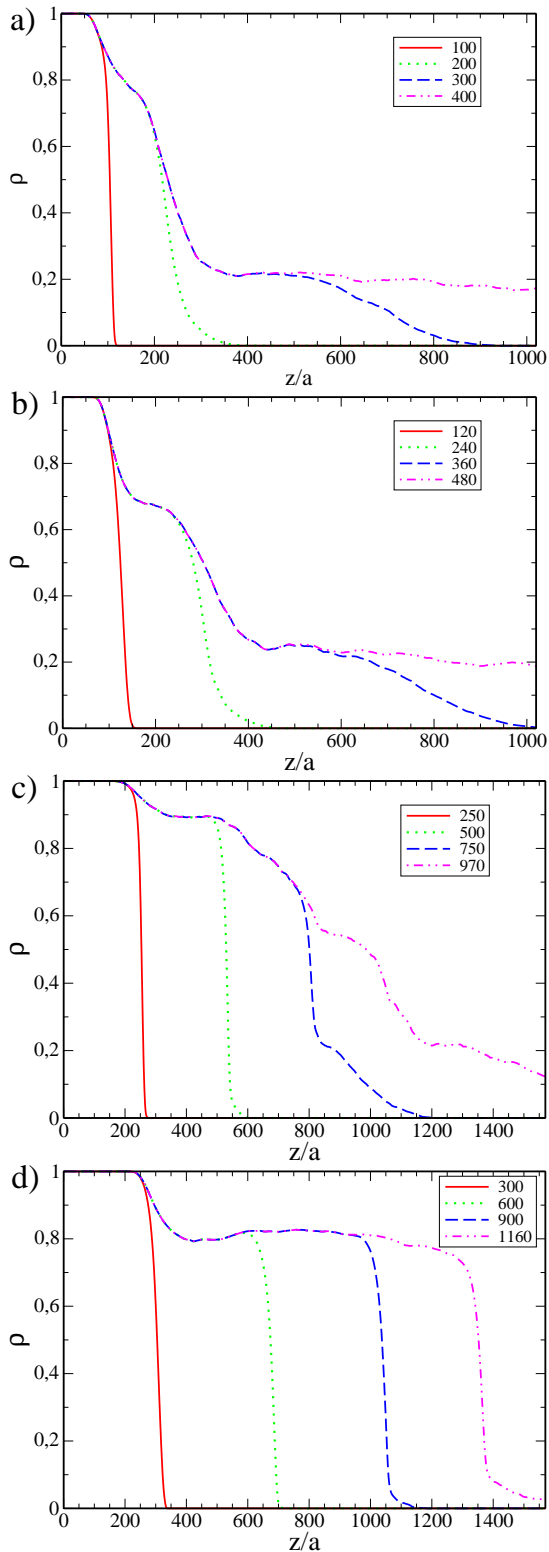

\includegraphics[clip,width=0.4\textwidth,angle=0]{profileD010000P0.05.eps}\\
\includegraphics[clip,width=0.4\textwidth,angle=0]{profileD010000P0.10.eps}\\
\includegraphics[clip,width=0.4\textwidth,angle=0]{profileD050000P0.05.eps}\\
\includegraphics[clip,width=0.4\textwidth,angle=0]{profileD050000P0.10.eps}
\caption{Density profiles of the films grown with:
(a) $G={10}^4$, $\epsilon=0.05$; (b) $G={10}^4$, $\epsilon=0.1$; (c) $G=5\times {10}^4$, $\epsilon=0.05$;
(d) $G=5\times {10}^4$, $\epsilon=0.1$.
In each plot, the reduced times $t/\tau$ (number of aggregated particles per substrate site) are indicated.
}
\label{density}
\end{figure}

Figs. \ref{density}a,b show the density profiles of films
deposited with $G={10}^4$ and two values of $\epsilon$ and
Figs. \ref{density}c,d show the density profiles of films
deposited with $G=5\times {10}^4$ and the same values of $\epsilon$.
The compact wetting layers correspond to $\rho\approx 1$ near the electrode.
At larger heights, $\rho\left( z\right) <1$ indicates that the film is porous;
the illustrations in Fig. \ref{sections} suggest that most pores are open,
but formation of isolated closed pores is also possible.

For $G={10}^4$ (Figs. \ref{density}a,b)
at short times ($t/\tau\lesssim 100$), the wetting layer is formed and a narrow region with
large density gradient is observed above that layer, which corresponds to a rough film surface.
As time increases, the wetting layer remains with approximately the same thickness:
$h\approx 70a$ for $\epsilon =0.05$, $h\approx 80a$ for $\epsilon =0.1$;
above that region, a plateau with $\rho\approx 0.6$ is formed for $\epsilon =0.1$
(from $h\approx 100a$ to $h\approx 350a$).
This feature is not observed for $\epsilon =0.05$ until the maximum times simulated here.
A region with large density gradient is then observed, which corresponds to the
nucleation and growth of branches above the compact layer.
At long times, the density for large $z$ continuously decreases, with a continuous increase
in the average film height.

Figs. \ref{density}c,d show the density profiles of films
deposited with $G=5\times {10}^4$ and two values of $\epsilon$.
The wetting layer is much thicker than that for $G={10}^4$,
but the effect of $\epsilon$ on its thickness is also small.
The density plateau formed above that
layer depends on $\epsilon$: $\rho\approx 0.9$ for $\epsilon=0.05$
[Fig. \ref{density}(c)], $\rho\approx 0.8$ for $\epsilon=0.1$ [Fig. \ref{density}(d)].
It corresponds to the thick initial branches separated by narrow gaps, as shown in
[Fig. \ref{sections}(b)].
For $\epsilon =0.05$, the continuous decrease of the density for large $z$ is
observed; for $\epsilon =0.1$, the plateau with $\rho\approx 0.8$ grows
until the maximal simulated times.

The long time density profiles clearly show the crossover from a compact film
($\rho\approx 1$) near the electrode to a film with large density gradient.
The region with large gradient may end at a density plateau, which is
followed by another region with large density gradient.
In all cases, the first of those regions corresponds to the onset of unstable
growth, in which branches begin to grow from the protuberances of the surface.
The thickness of these branches increases with $G$, as illustrated in
Figs. \ref{sections}(a) and \ref{sections}(b).

\begin{figure}[!ht]
\includegraphics[clip,width=0.5\textwidth,angle=0]{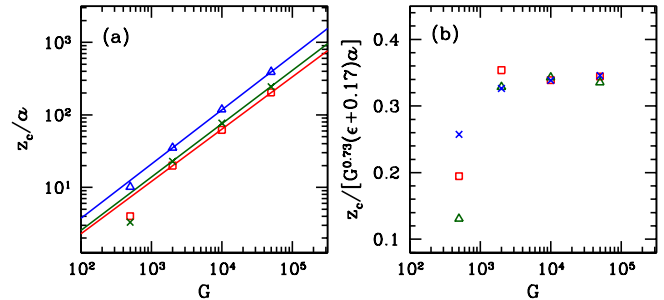}
\caption{(a) Crossover height as a function of the parameter $G$ for
$\epsilon =0.05$ (red squares), $\epsilon =0.1$ (green crosses), and $\epsilon =0.25$
(blue triangles).
Lines are least squares fits of each data set.
(b) Scaled crossover height as a function of $G$ for the same values of $\epsilon$.
}
\label{zc}
\end{figure}

For a given set of parameters, we define a crossover thickness $z_c$ as the position
$z$ in which the density measured at long times is $\rho_c=0.99$.
Figure \ref{zc}(a) shows $z_c$ as a function of $G$ for three values of $\epsilon$.
The effect of $\epsilon$ is small, but a significant variation with $G$ is observed.
The fits in Figure \ref{zc}(a) for each value of $\epsilon$ give
\begin{equation}
\frac{z_c}{a} \sim G^\gamma
\label{zcG}
\end{equation}
for large $G$, with $\gamma$ between $0.72$ and $0.75$.

In order to account for the effect of $G$ and $\epsilon$, Fig. \ref{zc}(b) shows a
scaled crossover thickness as a function of $G$.
It considers $\gamma =0.73$, which is the exponent that provides the best data collapse for
large $G$ and a correction factor depending on $\epsilon$.
The relation that follows from this scaling plot for large $G$ is
\begin{equation}
\frac{z_c}{a} \approx 0.34 G^{\gamma} \left( \epsilon +0.17\right) . 
\label{zcscaling}
\end{equation}

\subsection{Local roughness}
\label{localroughness}

\begin{figure}[!ht]
\includegraphics[clip,width=0.5\textwidth,angle=0]{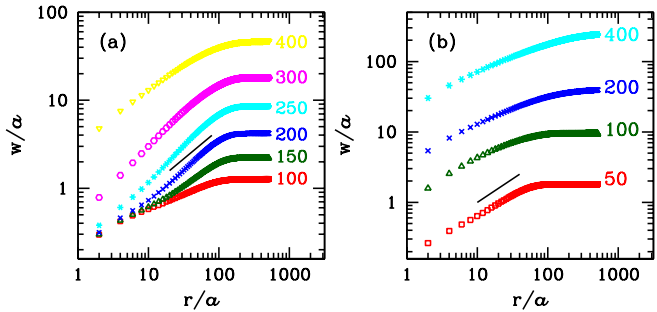}
\caption{Local roughness as a function of box size for the times $t/\tau$ indicated in the plots,
with: (a) $G=5\times {10}^4$, $\epsilon=0.10$; (b) $G=2\times {10}^3$, $\epsilon=0.05$.
The black lines in both plots have slope $2/3$ ($\approx\alpha_{VLDS}$).
}
\label{wlocal}
\end{figure}

In Figs. \ref{wlocal}a and  \ref{wlocal}b, we show the local roughness as a function
of box size at several times, for $\left( G=5\times {10}^4,\epsilon =0.1\right)$ and 
$\left( G=2\times {10}^3,\epsilon =0.05\right)$, respectively.

For $G=5\times {10}^4$ and $t/\tau\leq 100$, we obtain $w/a<1$ [Fig. \ref{wlocal}(a)],
which means that the film surface is almost flat, even  after deposition of several layers.
The scaling of $w$ with $r$ for intermediate times ($t/\tau=150$ and $200$ in
Fig. \ref{wlocal}(a) is consistent with normal scaling.
The slope for small $r$ is near $0.67$, which is the roughness exponent of the VLDS class.

For longer times ($t/\tau\geq 250$), the splitting of the curves for small $r$ is typical
of anomalous scaling.
However, for a fixed small $r$, the roughness increases faster than a power law, thus the
exponent $\kappa$ [Eq. (\ref{fvanomalous})] cannot be estimated.
This anomaly is only a signature of the onset of unstable growth.

For $G=2\times {10}^3$, the local roughness shows the typical features of anomalous
scaling at short times [Fig. \ref{wlocal}(b)].
Again, this behavior is not representative of a true anomalous scaling, but a consequence of the
rapid growth of branches separated by deep gaps.
At $t/\tau =50$, the slope of the $\log{w}\times\log{r}$ plot is also near the VLDS value $0.67$.

\subsection{Correlation length}
\label{correlationlength}

For small values of $G$ (e. g. $G=100$), the autocorrelation function
$\Gamma$ has only a shallow minimum for short times,
and oscillations disappear at $t/\tau \sim 50$ and longer.
This occurs because the instability develops at short times.

\begin{figure}[!ht]
\includegraphics[clip,width=0.5\textwidth,angle=0]{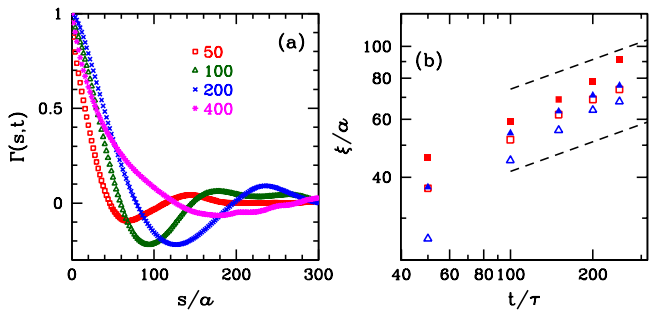}
\caption{(a) Autocorrelation function as a function of the horizontal distance $s/a$ at times $t/\tau$
indicated in the plot, for $G=5\times 10^4$ and $\epsilon=0.05$.
(b) Correlation length as a function of time for
$\left( G=5\times {10}^4,\epsilon =0.05\right)$ (squares)
$\left( G=5\times {10}^4,\epsilon =0.1\right)$ (triangles),
calculated with $k=0$ (filled symbols) and $k=0.1$ (empty symbols).
The black dashed lines have slope $0.3$ ($\approx \nu_{VLDS}$).
}
\label{cor}
\end{figure}

For $G=5\times {10}^4$ and $\epsilon =0.05$, the
autocorrelation function is shown in Fig. \ref{cor}(a).
As time increases, the first minimum of $\Gamma$ is enhanced (i. e. becomes deeper), which
indicates a rapid growth of the protuberant parts of the film.
For $t/\tau >200$, the depth of the minimum is reduced, which indicates that a characteristic
length of height fluctuations does not exist in this regime.

The time evolution of the correlation length $\xi$ for
($G=5\times {10}^4$, $\epsilon =0.05$) and
($G=5\times {10}^4$, $\epsilon =0.1$) is shown in Fig. \ref{cor}(b),
considering Eq. (\ref{defxi1}) with $k=0$ and $k=0.1$.
The plotted data is restricted to times $t/\tau \leq 250$, in which there is no evidence of
unstable growth.
Each data set shows a different evolution and the restricted time range does not allow
a reliable extrapolation.
However, the average trend of those data is consistent with the exponent $\nu\approx 0.30$
of the VLDS class, as shown in Fig. \ref{cor}(b).

\section{Relation with other models of compact film growth}
\label{relation}

The simulation results in Secs. \ref{localroughness} and \ref{correlationlength}
suggest that the roughening at short times has VLDS scaling.
For this reason, here we analyze the relations with similar models with the same scaling.

The particle diffusion of our model is similar to that of a model
introduced in Ref. \protect\cite{CDLM}, which is hereafter called lateral aggregation of
diffusing particles (LADP).
In the LADP, each particle is released above the deposit at a randomly chosen position
$\left( x,y\right)$ and follows a vertical trajectory until reaching the top of the column
of the deposit.
The particle subsequently executes random hops to the top of NN columns and permanently
aggregates when it has a lateral NN or after executing $G$ hops.
Reference \protect\cite{CDLM} also showed that the LADP has VLDS roughening.

There are some differences between our electrodeposition model and the LADP:
here we consider a diffusive motion of the cations in solution (instead of the collimated flux
of LADP); the hops of mobile particles are allowed to any spatial direction, which may lead
to overhang formation (while LADP films are compact); the particle may move if it has two
or more occupied NNs (while the LADP corresponds to $\epsilon=0$).
However, these differences are not important at short times, in which
the electrodeposition model produces almost compact deposits with small
surface roughness, consequently with no significant protuberance.
Moreover, the parameter $\epsilon>0$ has weak effects on the morphology of the compact deposits;
the main parameter affecting the surface morphology is $G$, which has the same role as in the LADP.
For the above reasons, the compact film morphology at short times in our electrodeposition model
is expected to be similar to that in the LADP.
This sets a connection between those models and provides additional evidence for the VLDS scaling
of our electrodeposition model at short times.

The scaling of the correlation length of the LADP is of particular importance here \cite{CDLM}:
\begin{equation}
\xi \sim A G^{1/2} {\left(\frac{t}{\tau}\right) }^{\nu} a ,
\label{xiCDLM}
\end{equation}
with $\nu\approx 0.3$ and $A\approx 0.5$.
In the electrodeposition model, the amplitude $A$ has a reduction by a factor
$\sim 8$, which is probably due to the frequent rejection of hop attempts
(the parameter $G$ in LADP refers only to hops that can be executed).

The LADP was designed for producing deposits with the same morphology of
those of the CV model, which was introduced in Ref. \protect\cite{cv}.
The CV model describes collective adatom diffusion during the deposition and is frequently
used to describe molecular beam epitaxy~\cite{etb}; however,
applications to other deposition processes were already proposed \cite{tanyakarn,dungreis2015}.
The atomic flux in the CV model is also collimated, with a flux $F$ defined as the
number of incident atoms per site per unit time.
The solid-on-solid condition is assumed.
In one time unit, each surface atom executes an average of $D\epsilon^{n-1}$ hops to NN columns,
where $D$ is the terrace diffusion coefficient given in Eq. (\ref{defD}) and $\epsilon$ is
defined as in Eq. (\ref{Eb}).
The main parameter to determine the surface features in the CV model is the dimensionless ratio
\begin{equation}
R\equiv \frac{D}{F} =\frac{h_0}{F} \exp{\left( -\frac{E_s}{k_BT}\right)} .
\label{defR}
\end{equation}

When $\epsilon = 0$ (i. e. irreversible lateral aggregation),
Ref. \protect\cite{CDLM} showed that the CV model produces deposits
with the same roughness of the LADP model with
\begin{equation}
G \sim R^{2\nu} .
\label{Geq}
\end{equation}
Recently, Ref. \protect\cite{cv2015} showed that $\epsilon$ has a very small effect
on the surface features of CV deposits for $0\leq\epsilon\leq 0.25$.
Thus, the LADP also produces film morphology similar to that of the CV model for $\epsilon >0$
if their parameters are related by Eq. (\ref{Geq}).

The above discussion leads to a connection between our electrodeposition model at short times
and the CV model.
The mobile particle diffusion of our model may be interpreted as an
approximation of a collective particle diffusion process.
Equations (\ref{defR}) and (\ref{Geq}) show that $G$ increases with the diffusion coefficient
$D$ and decreases with the deposition rate $F$, as discussed in Sec. \ref{interpretation}:
\begin{equation}
G \sim {\left( \frac{h_0}{F}\right)}^{2\nu} \exp{\left( -\frac{2\nu E_s}{k_BT}\right)} .
\label{GEs}
\end{equation}
Equation (\ref{GEs}) also confirms the Arrhenius form expected for $G$, which was anticipated
in Sec. \ref{interpretation}.

This connection with VLDS models at short times is the basis for explaining the formation of
the films with layers of different morphology when the particle flux is diffusive and to
describe the effect of physico-chemical parameters on the crossover thickness.

\section{The crossover from compact to branched films}
\label{crossover}

\subsection{Scaling approach}
\label{scaling}

\begin{figure}[!ht]
\includegraphics[clip,width=0.4\textwidth,angle=0]{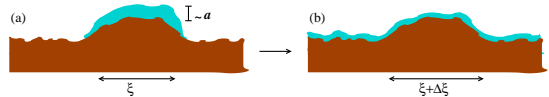}
\caption{(a) Schematic view of a deposit (brown) with a protuberant part
and a layer of particles (light blue) formed after cation reduction at that protuberance.
The correlation length $\xi$ is of the order of the lateral size of the protuberance.
(b) Spread of the new layer by diffusion, with increase of the protuberance size.
}
\label{scheme}
\end{figure}

Consider the scheme of Fig. \ref{scheme},
in which a large fluctuation (a hill) appears in the film surface.
The autocorrelation function oscillates and has a pronounced minimum,
as illustrated in Fig. \ref{cor}(a), showing that the lateral size of that fluctuation is of
the same order of the correlation length $\xi$ of the initial kinetic roughening.

Most of the cations flowing to the region shown in
Fig. \ref{scheme}(a) are reduced in the protuberant part of the surface.
Thus, most mobile particles are formed at that region, while a small fraction
is formed in the valleys at the hill sides.
In a time interval $\tau$, a layer of particles with thickness of order $a$ is formed,
thus the velocity of vertical growth of the protuberance is of order
\begin{equation}
v \sim \frac{a}{\tau} . 
\label{defv}
\end{equation}
Of course this relation omits a numerical factor ($>1$) at the right side
due to the preferential reduction at the protuberant part of the surface.
However, this factor is of order $1$ because we assume an initially small fluctuation.

We expect that the accumulated layer of particles diffuses to the neighboring
valleys, as shown in Fig. \ref{scheme}(b).
This leads to an increase of the correlation length $\xi$.
The velocity $u$ of lateral flow of material is
\begin{equation}
u \sim \frac{d\xi}{dt} \sim A G^{1/2} \frac{a}{\tau} {\left(\frac{t}{\tau}\right) }^{\nu -1} ,
\label{defu}
\end{equation}
where Eq. (\ref{xiCDLM}) was used.
The time increase of $\xi$ is always slower than linear because $\nu\leq 1$
(dynamical exponent larger than $1$).
For this reason, $u$ decreases in time.

If $u>v$, then surface diffusion is able to balance the effect of the preferential
adsorption at the protuberance.
Thus, normal kinetic roughening is expected, i. e. roughening dominated by the
surface diffusion (VLDS scaling).
This is expected to occur at short times.
On the other hand, if $u<v$, the mass accumulated at the protuberance moves slowly
to the neighboring valleys, thus the protuberance height increases.
This leads to  the unstable growth and is expected at long times.
A crossover between the two regimes is expected when those velocities match: 
\begin{equation}
v\approx u \qquad \left( t=t_c\right) .
\label{condcrossover}
\end{equation}

Using Eqs. (\ref{defv}), (\ref{defu}), and (\ref{condcrossover}), we obtain
\begin{equation}
\frac{t_c}{\tau} \sim G^\gamma ,
\label{scalingtc}
\end{equation}
with
\begin{equation}
\gamma = \frac{1}{2\left( 1-\nu\right)} .
\label{gamma}
\end{equation}
The corresponding crossover thickness is
\begin{equation}
z_c \sim v t_c \sim a G^\gamma .
\label{zcscalingtheory}
\end{equation}
Using the VLDS exponent $\nu\approx 0.30$ \cite{crsosreis}, we obtain
$\gamma \approx 5/7 \approx 0.71$.
This agrees with Eq. (\ref{zcG}), with an
exponent $\gamma$ in excellent agreement with the simulation results  presented
in Sec. \ref{densityprofile}.

This approach is proposed for growth on a flat substrate, but small changes in the
crossover thickness $z_c$ are expected if the substrate is rough.
For large values of $G$ and substrate roughness of $1$-$10$ lattice units (which
typically corresponds to a few nanometers), an initial smoothening process is shown in
Ref. \protect\cite{smoothening} for deposition with similar models with collective surface
diffusion.
The characteristic film thickness necessary for smoothening is
$z_s\sim R^{-0.4}{\left( \xi_i /a\right)}^\nu$, where $\xi_i$ is the correlation length of
the initial pattern; using Eq. (\ref{Geq}), we obtain an estimate $z_s\sim G^{-0.67}{\xi_i}^\nu$
for our model.
For $G\sim {10}^3$ and a large initial correlation length $\xi_i /a\sim {10}^3$, it gives $z_s<1$;
for larger $G$, $z_s$ is smaller, thus it is negligible in comparison with $z_c$.
The physical interpretation of this result is that deposition with large $G$ (or $R$) rapidly
fills the valleys of the initial rough pattern, which consequently do not affect the crossover
to the branched morphology.

\subsection{Effect of activation energy and temperature}
\label{activationenergy}

Now we consider the relations with CV models discussed in Sec. \ref{relation} to determine the
crossover thickness $z_c$ as a function of parameters that may be measured experimentally.
As a first approximation, we use $h_0\sim {10}^{13}$s${}^{-1}$ in Eq. (\ref{GEs}),
which is the value considered when the solid film grows from vapor \cite{etb}
(although we understand that there are significant difference when the solid surface is in
contact with a solution).
Using Eq. (\ref{zcscaling}) for very small $\epsilon$ and Eq. (\ref{gamma}), we obtain
\begin{equation}
\frac{z_c}{a} \sim \frac{{10}^4}{F^\lambda}\exp{\left( -\frac{\lambda E_s}{k_BT}\right)} ,
\label{zcexp}
\end{equation}
where
\begin{equation}
\lambda = 2\nu\gamma = \frac{\nu}{1-\nu} \approx 0.43 .
\label{lambda}
\end{equation}
The parameter $\epsilon >0$ has weak effect in Eq. (\ref{zcexp}).
The amplitude in that equation is a rough estimate due to the uncertainty in $h_0$
and due to the rejection of hop attempts in the electrodeposition model, which also affects the
amplitudes in equations such as (\ref{GEs}), as discussed in Sec. \ref{relation}.

The present model is suitable for a fully supported electrolyte, in which cations in solution move
with no bias to the electrode.
Moreover, no effect of additives or other mechanisms that hinder unstable growth is
considered.
If diffusion bias, additives, etc are present, longer stable regimes are expected, corresponding
to larger values of $z_c$. 
For this reason, we understand that the present estimate of $z_c$ is a lower bound for the
thickness of a compact layer that can be obtained in electrochemical deposition of a
given material.
This lower bound may be useful in works whose aim is only to obtain branched morphology; these films
may be of interest as catalysts or for their particular wetting properties;
see, e. g. Ref. \protect\cite{talebJMCA2013} for cobalt film electrodeposition.

\subsection{Possible applications}
\label{relevance}

Recent works on electrodeposition of several materials show evidence that surface diffusion
is an important mechanism to determine the large scale morphology
\cite{grunder,PB,aryanfar,brandt2015,desai,zimmer,guoEA2010,tanyakarn,mamme}.
For instance, Xu et al \cite{xu} analyzed the kinetic roughening of electrodeposited NiP films
and showed exponents very close to those of the VLDS class:
$\alpha = 0.70$ and $\beta = 0.16$ (to be compared with $\alpha_{VLDS}\approx 0.67$
and $\beta_{VLDS}\approx 0.20$ \cite{crsosreis,carrasco2016}).
These results give additional support to the model presented in this work.

The formation of a compact layer before the growth of a branched structure is observed in
electrochemical deposition of some metals, such as zinc \cite{argoul1988},
copper \cite{kahanda,pastor,lafouresse,debruyn,legerPRE1998}, and silver \cite{pasquale}.

In Ref. \protect\cite{kahanda}, the thickness of the compact copper layer increases
as the overpotential becomes less negative, corresponding to a decrease in the average
growth rate.
The thicknesses of the compact layers were not reported in that work, but inspection of
their images shows an increase by a factor of approximately $2$ when
the growth rate decreases by a factor of approximately $3$, with
change of overpotential from $-0.35$V to $-0.15$V.
This result is qualitatively consistent with the dependence of $z_c$ on $F$
in Eq. (\ref{zcexp}) with $\lambda<1$ [Eq. (\ref{lambda})].

The FCC structure of copper differs from the simple cubic lattice used in our simulations,
thus our results cannot provide a quantitative description of that material.
However, it is interesting to estimate the order of magnitude of quantities that follow
from the model application.
Electrodeposition of Cu is frequently reported with compact films with thicknesses of a
few micrometers, thus we consider the case of $z_c\sim 1\mu$m at room temperature.
We also consider a lattice constant $a\approx 0.3$nm and a current $j\sim 10$mA/cm${}^2$,
consistently with Refs. \protect\cite{kahanda,lafouresse}; this gives $F\sim 30$monolayers/s.
Substitution in Eq. (\ref{zcexp}) gives an exponential factor
$\exp{\left[ -\lambda E_s/\left( k_B T\right)\right]}$ of order $1$, which means that
$E_s\lesssim 0.05$eV.
This is a very small activation energy, so that diffusion lengths of the deposited
atoms can be very large, which highlights the importance of adsorbate diffusion in
electrodeposition.
However, due to the drastic approximations involved in the derivation of Eq. (\ref{zcexp}),
an accurate estimate of $E_s$ cannot be obtained.

Several technological applications such as photovoltaic, catalysis, sensing, and batteries
require highly porous thin layers with limited compact layer, which justifies the present
study.

For dye-sensitized solar cell applications, high specific surface of porous
materials offers the possibility of high dye loading to sensitize the semiconductor
material. Additionally, the branch or the grain size allows the light scattering
within the material, which results in extending the path of light, and in turn
increases the probability of photons being captured by the sensitizers.
However, the increase of film thickness if necessary for the formation of large branches
or grains, which causes dye loading reduction and the enhancement of charge
recombination processes. The balance between all these conflicting properties
requires an accurate tuning of the film thickness \cite{hafed03}.
For instance, for porous films made of ZnO aggregates, the optimum thickness to
obtain this balance is about $10\mu$m \cite{hafed11}.

For applications such as batteries \cite{hafed12},
sensors \cite{hafed13,hafed14}, or catalysis \cite{hafed15},
porous materials provide high active surface areas
and shortened pathways for the fast diffusion of reactive species in solution
toward the surface. For catalysis and sensors, different optimal thicknesses were
reported in the literature, ranging from a few hundred nanometers to a few tens of
micrometers, depending on the material, the type of sensor, and the targeted species.
However, as the thin film thickness increases, so does the surface area,
but the diffusion path length within the material becomes longer, which
counterbalances the beneficial effect of the surface area increase. For
optimization, a balance between the surface area and the diffusion path length
must be found \cite{hafed16,hafed17}.

Other model features may also be considered for some applications.
For instance, in metal electrodeposition, Ustarroz et al \cite{ustarroz,desai} observed that
the initial stages of growth are dominated by coarsening of islands whose building blocks
are clusters of a few nanometers, instead of single adatoms.
Their sizes may be important for a correct estimate of the smallest lengthscale $a$ and affect the
interpretation of activation  energies.
It is also possible that roughening of a compact film is not described by the VLDS equation;
for instance, electrodeposited Ni films of Ref. \protect\cite{brandt2015} and Prussian blue
films of Ref. \protect\cite{PB} have kinetic roughening exponents of the Mullins-Herring
equation \cite{mh}, which is the diffusion-dominated growth model based on Eq. (\ref{vlds})
with $\lambda_4=0$.
In these cases, the scaling exponents of Sec. \ref{scaling} should be changed to consider
$\nu=0.25$ of that class.

\subsection{Relation to other models}
\label{relationmodels}

Several works have already shown the formation of thin compact films in growth models with
diffusion-limited aggregation of the incident species.
However, in most cases, one of the following mechanisms was present: large concentrations
of cations in solution, which reduces the thickness of the diffusion layer near the
film surface and allows a more uniform flux along the surface of the deposit, or
biased diffusion of cations, which also has the asymptotic effect of homogenizing
the adsorption rate along the surface.

A simple example was provided in Ref. \protect\cite{uwaha} with DLA of particles that move
in a solution with fixed concentration: for large concentrations,
dense films with surfaces not very rough were obtained; for intermediate concentrations,
porous films resembling ballistic deposits \cite{barabasi} were obtained; for low
concentrations, an initial fractal structure is observed, but there is a crossover to growth
of a film with finite porosity at long times.

Another example is multiparticle biased DLA (MBDLA) \cite{sanchez,castro1998},
in which cations execute random walks biased toward the electrode ($-z$ direction).
In Ref. \protect\cite{castro2000}, MBDLA was studied in two dimensions (one-dimensional
substrate) including collective diffusion of the adsorbed atoms.
The formation of thick branches above a compact film was observed when the surface
diffusion coefficient was one order of magnitude smaller than that of diffusion in
solution and the bias was low; if those coefficients were equal, then a thick
compact film was obtained during all the simulated time.
The bias in the particle movement represents the effect of the electric field in solution
and always suppresses the unstable growth at long times.

Reference \protect\cite{nielsen} recently presented a continuum electrodeposition model
with extended space-charge regions which, for small applied potential, also shows formation
of a compact film near the electrode, with thickness increasing with the cation
concentration.
The phase-field models for electrochemical deposition proposed in Refs.
\cite{nicoliJSTAT2009,cogswell} also produce deposits with a compact layer near the (one-dimensional)
electrode.
In Ref. \protect\cite{nicoliJSTAT2009}, the main interest was the comparison of the branched morphology
with that of experiments on copper \cite{kahanda,debruyn,legerPRE1998} instead of the transition
discussed in this work.

If a small bias in the motion of cations is included in our model, we also expect that the
unstable growth will be asymptotically suppressed. 
However, the present approach for the compact-to-branched crossover remains valid if
this suppression takes place with a sufficiently large film thickness $z_s$, i. e. if $z_s\ll z_c$.

\section{Conclusion}
\label{conclusion}

We introduced an electrodeposition model in a simple cubic lattice in which diffusion-limited
aggregation of cations in a solution is followed by their reduction and surface diffusion of
the adsorbed species, with a maximum of $G$ hops to nearest neighbor sites.
Films with overhangs and pores may be formed because adsorption occurs at the first point in
which the cation is in contact with the deposit and because diffusion to all directions is
allowed, under the condition that all adsorbed particles remain connected.
At short times, compact films with smooth surfaces are formed because diffusion
favors aggregation to sites with large number of nearest neighbors.
In this regime, the surface roughness and the correlation length scale with exponents close to
those of the VLDS class; their scaling is consequently approximated by that of a previously
studied model of irreversible lateral aggregation or, equivalently, the Clarke-Vvedensky model
of deposition and diffusion without detachment from lateral neighbors.
At long times, the films have branches whose widths increase with the number of allowed
diffusion hops.

The presence of dense and branched layers in a single film is suitable for applications that
require multifunctional behavior.
That feature was actually observed in electrodeposited films of several materials.
This motivated the study of the relation between the maximal thickness of the compact layer,
$z_c$, and the parameter $G$, which led to a relation with the activation energy and temperature in
an equivalent growth process with collective adatom diffusion.
A scaling approach showed that, at short times, the particles formed at surface protuberances
can diffuse to distant points, homogenizing the growth rate along the surface; however, at
long times, the lateral size of those protuberances (correlation length) is large, thus
the amplitude of surface fluctuations grows in time and branches are subsequently formed.
The predictions of that approach were confirmed by numerical simulations.

Whether one seeks multilayers with different features or aims at promoting a
layer with specific features, understanding the structure and formation of these
layers by electrodeposition is of crucial importance.
To this aim, in this paper, we have focused on the onset of the instability
related to the transition from dense to ramified layer morphologies.
This understanding should allow better control on the thickness of different
parts of the layer to adjust to various applications.

\section*{Acknowledgment}

F.D.A. Aar\~ao Reis acknowledges support from CNPq and FAPERJ (Brazilian agencies).


\end{document}